\documentstyle[multicol,epsfig,aps]{revtex}
  
\begin{document}

\title{Clustering dynamics of Lagrangian tracers in 
free-surface flows}
\author{J\"org Schumacher$^{1,2}$ and Bruno Eckhardt$^2$}
\address{$^1$ Department of Mechanical Engineering, Yale University, 
New Haven, Connecticut 06520-8284, USA\\
         $^2$ Fachbereich Physik, Philipps-Universit\"at Marburg,  
D-35032 Marburg, Germany}  
\date{\today}  
\maketitle 

\begin{abstract}
We study the formation of clusters of passive Lagrangian tracers in 
a non-smooth turbulent
flow in a flat free-slip surface as a model for particle dynamics on free
surfaces.  Single particle and pair dispersion show different behavior for short
and large times:  on short times particles cluster exponentially rapidly until
patches of the size of the divergence correlation length are depleted; on
larger times the pair dispersion is dominated by almost ballistic 
hopping between
clusters.  We also find that the distribution of particle density is close
to algebraic and can trace this back to the
exponential distribution of the divergence field of the surface flow.

\end{abstract}

\vspace{0.5cm}
\begin{multicols}{2}
The Lagrangian evolution of passive tracers in turbulent flows has 
attracted considerable attention
recently. Particle tracking techniques have been developed that
allow for a detailed observation of their motion 
even in high-Reynolds number turbulence \cite{OM,Porta,Pinton}.
The statistics of few particle clusters has been used to obtain
information on the Lagrangian statistics of the flow field
\cite{Pumir}. Simplified models of passive scalars evolution,
e.g. the Kraichnan model with its delta-correlated random 
velocity fields, have provided important insights into the 
origin of intermittency corrections to scaling laws
\cite{Kraichnan}. The advection of particles that are not
neutrally buoyant gives rise to clustering
and this process has been suggested to be essential for the 
formation of rain\cite{Balkovsky01}. Similar clustering phenomena
should appear for bubbles or inertial particles in turbulent flows 
\cite{Haarlem}.
The problem we consider here is the advection of Lagrangian particles
in a flat free surface above a turbulent volume flow.
Previous approaches to the particle advection in such (compressible) flows
include random maps \cite{Ott} and
Kraichnan models 
with prescribed smooth \cite{Saich97} and non-smooth 
\cite{Gawedzki} spatial variations. Realistic flows 
have some finite time correlations, but, as we will demonstrate here,
they show further differences:
the distribution of values for the divergence of the flow field
is exponential (and not Gaussian \cite{Saich97}), the distribution of density
is algebraic (and not lognormal \cite{Saich97}) and the 
two-particle dispersion shows an almost ballistic regime for
large separations.
Furthermore, this behavior is outside the range of 
Kraichnan type models since the ratio between divergence and
velocity gradient fluctuations is such that the surface flow
belongs to a marginal situation where the Kraichnan
models predict neither clustering nor exponential separation
of particles \cite{Kraichnan,Gol01,Eck01}.

The experimental realization, the dynamics and the properties of the flat free
surface flows that we consider have been discussed in detail before
\cite{Gol01,Eck01}.  What is needed here is the presence of a $r^{2/3}$-scaling
in the inertial subrange due to connection to bulk turbulence in the volume
below, i.e. a non-smooth flow with finite time correlations which goes beyond 
all previous analytical approaches.
On the numerical side we integrate particle trajectories using a bicubic
spline interpolation which was checked by comparison with analytical examples
\cite{Yeung} and with a direct spectral evaluation of the velocity between the
grid meshs \cite{Rovelstad}.  The pseudospectral simulations of the
Navier-Stokes flow are based on grids
with $256\times 256\times65$ nodes, with a forcing that maintains a fixed energy
injection rate $\epsilon$ \cite{Eggers}.  A Taylor-Reynolds number
$R_{\lambda}=u^2_{x,rms}/[\nu (\partial_x u_x)_{rms}]=145$ was achieved and the
Kolmogorov length $\eta=(\nu^3/\epsilon)^{1/4}$ is 0.8 grid spacings.  The
Kolmogorov time is $\tau_{\eta}=(\nu/\epsilon)^{1/2}$.

\begin{figure}
\begin{center}
\epsfig{file=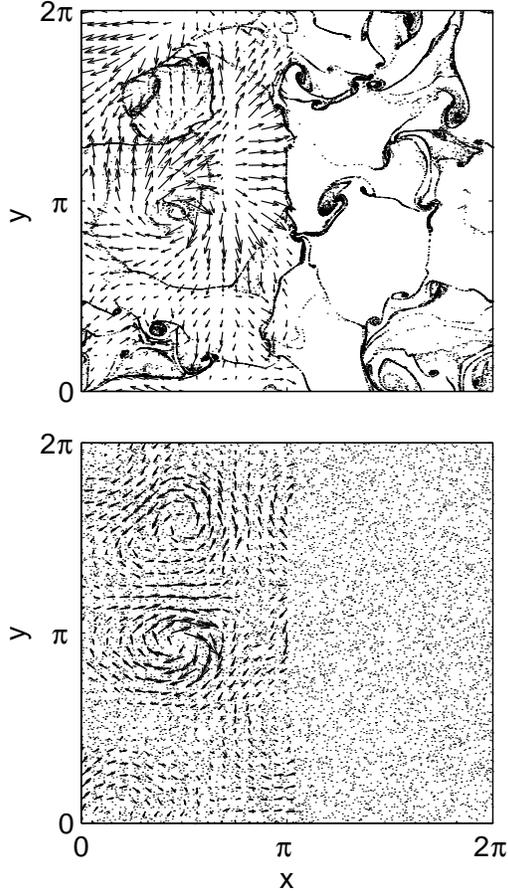,width=7cm}
\end{center}
\caption[]{Distribution of 36000 tracers and the instantaneous velocity field.
Upper panel:  full flow ${\bf v}$.  Lower panel:  solenoidal flow ${\bf v}_s$.
The snapshots for both cases were taken at $t/\tau_{\eta}=21$ after the start.
In order to highlight the tracer patterns the underlying flow fields are shown
in one half of the box.}
\label{fig1}
\end{figure}

A typical particle distribution that emerges from a uniform initial distribution
is shown in Fig.~\ref{fig1} (upper panel).  The particle dynamics shows two time
regimes, a quick clustering into elongated structures, followed by a slower
exchange of particles between structures.  Superimposed on the particle
distribution we show the surface flow that can be considered as a superposition
of an irrotational and a gradient part by the Helmholtz decomposition theorem,
\begin{equation}
{\bf v}={\bf v}_s+ {\bf v}_p =
{\bf\nabla\times}\phi(x,y)\,{\bf e}_z+{\bf\nabla}\psi(x,y)\,,
\end{equation}
with scalar potentials $\phi(x,y)$ and $\psi(x,y)$. 
It suggests that the aggregate of particles is dominated by the gradient
field, with the particles clustering in the minima of the 
potential $\psi$, similar to compressible Kraichnan flows \cite{Kraichnan}.
When the velocity field is projected onto
the solenoidal part ${\bf v}_s$ the particle distribution 
remains essentially uniform and there is no clustering
(lower panel of Fig.~1).

\begin{figure}
\begin{center}
\epsfig{file=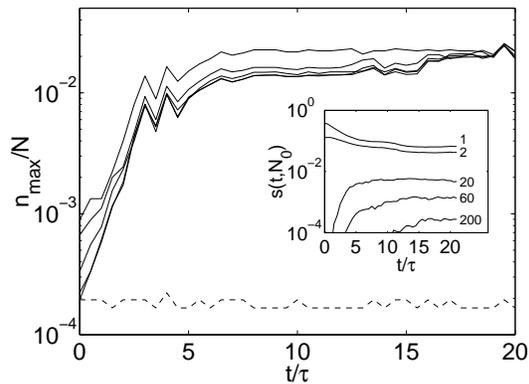,width=7cm}
\end{center}
\caption[]{Maximum particle number per cell $n_{max}/N$ vs.  $t/\tau$ for
particle numbers 36000, 18000, 9000, 4500, and 2250 (from bottom to top).
Dashed line is for ${\bf v_s}$ and solid lines are for ${\bf v}$.  The inset
shows the temporal evolution of the integrated probability $s(t,N_0)$ with
values of $N_0$ to the right.}
\label{fig2}
\end{figure}

The dynamics in this initial period is dominated by the exponential
contraction in ${\bf v}_p$. This follows, e.g. from 
the advection-diffusion equation for a smooth density $\rho$ with
diffusivity $D$,
\begin{equation}
\partial_t \rho =-({\bf \nabla}\cdot {\bf v}) \rho -({\bf v}\cdot\nabla)\rho + D
\Delta \rho\,,
\label{rhoeq}
\end{equation}
where the divergence patches cause an exponential variation that typically is
faster than the variations of the other terms \cite{Eck01}.  
The natural time scale
$\tau=\langle ({\bf\nabla\cdot v})^2\rangle^{-1/2}\approx 3.5\tau_{\eta}$ is 
approximately the lifetime of a divergence patch which comes out to be about
$5\tau_{\eta}$ (half width half maximum of the temporal divergence correlation
function). As expected, the divergence-free advection 
contributes little to the clustering.
For the discrete particles we measure the density by coarse graining, i.e.
dividing the plane into $256\times 256$ grid cells and count the particles
inside the cells.  The maximal number $n_{max}/N$ with total particle number,
$N$, increases initially exponentially, as demonstrated in Fig.~\ref{fig2}.  For
$t>\tau$ the exponentially fast formation crosses over to a slower regime.  The
accumulation of more particles into larger clusters continues:  the inset of
Fig.~\ref{fig2} shows the integrated probability to find cells with no fewer
than $N_0$ particles, $s(t, N_0)=\sum_{n=N_0}^N p(t,n)$, for different values of
$N_0$ with $1\le N_0\le N$. The probabilities continue to vary even when 
the maxima remain essentially constant.

\begin{figure}
\begin{center}
\epsfig{file=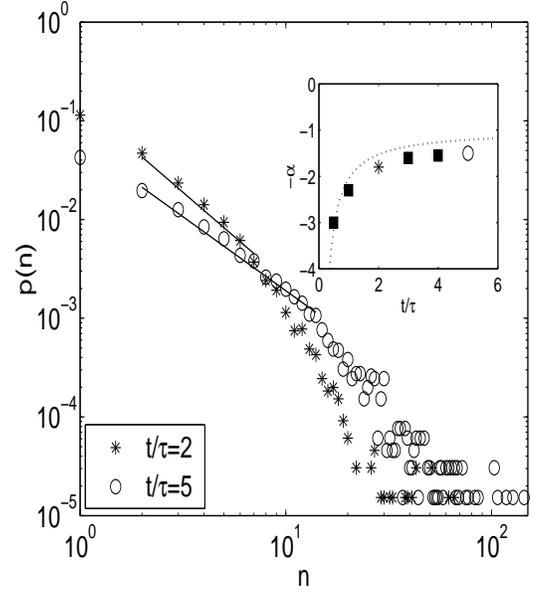,width=7cm,height=8cm}
\end{center}
\caption[]{Probability density $p(n)$ for different times during the cluster
evolution. Straight lines indicate fits with an algebraic law. 
The inset shows
the exponents $\alpha$ of $p(n)\sim n^{-\alpha}$ as a function of time $t/\tau$.
The dotted line follows from (4).}
\label{fig3}
\end{figure}

\begin{figure}
\begin{center}
\epsfig{file=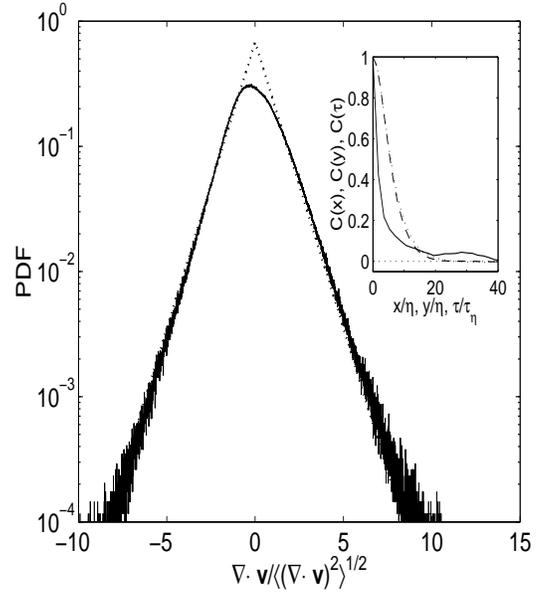,width=7cm,height=8cm}
\end{center}
\caption[]{PDF of the divergence of the surface flow ${\bf v}$.  The dotted line
is an exponential fit, $p(\lambda)=0.68\exp(-|\lambda|/0.95)$ with
$\lambda=({\bf\nabla\cdot v}) \tau$.  The inset shows the spatial and temporal
correlation functions of the divergence field, $C(x)$ over $x/\eta$, $C(y)$ over
$y/\eta$, and $C(\tau)$ over $\tau/\tau_{\eta}$.}
\label{fig4}
\end{figure}

The divergence also determines the particle distribution $p(n)$, 
as shown in Fig.~\ref{fig3}. In Kraichnan type models this distribution
comes out to be lognormal for smooth flows \cite{Saich97,Kraichnan} and
delta-like for non-smooth cases \cite{Gawedzki}. 
Although we were limited to moderate resolutions and particle numbers, 
our data in 
Fig.~\ref{fig3} are closer to 
an algebraic
distribution which is consistent with the 
exponential distribution of the divergence (see Fig.~\ref{fig4}). 
Neglecting
the diffusion term in (\ref{rhoeq}), the density increases
exponentially in the Lagrangian frame, 
$\gamma(t)=\rho(t)/\rho(0) = \exp(-\int_0^t({\bf\nabla\cdot v}) dt^{\prime})$. 
For $t\leq 1$ we can simplify the exponent to $\lambda t$
where
$\lambda = {\bf\nabla \cdot v}$ (in units of $\tau^{-1}$). 
If we assume that the density 
variations are faster than the changes in the velocity field, we
have local fluctuations in the divergence that give rise to 
locally varying density fluctuations. If $P(\lambda)$ is the 
probability density function (PDF) for the divergence, 
then the PDF for $\gamma$ becomes
\begin{equation}
\tilde{P}(\gamma)=\int d\lambda\;\delta(\gamma - \mbox{e}^{\lambda t}) P(\lambda)
=\frac{1}{\gamma t} P(\ln \gamma/t)\,.
\label{psigmaeq}
\end{equation}
Hence, if the divergence fluctuations are Gaussian, as in smooth 
Kraichnan flows \cite{Saich97}, the PDF of the density fluctuations 
is lognormal.
However, in the case of the surface flow the divergence fluctuations
have a filamentary small-scale structure \cite{Eck01} (refered
to as a shocklet (negative divergence) in compressible 
supersonic turbulence \cite{Porter98}).
The small scale structures appear in the PDF as exponential 
tails (see Fig.~\ref{fig4}).
If we let $P(\lambda) = 1/(2s)\mbox{e}^{-|\lambda|/s}$, where
$s\approx 0.95$ from Fig.~4, then
\begin{equation}
P(\gamma)\sim |\gamma|^{-1-1/(st)} \,.
\label{psigmaeq1}
\end{equation}
As shown in the inset of Fig.~\ref{fig3}, 
the slope $-\alpha$ in the tails of the distribution decreases with time, from
about $-3.0$ at $t=0.5$ to $-2.3$ 
at $t=1$, in good agreement with the prediction from (\ref{psigmaeq1}).
For much longer times the discretness of the particles
shows up and the distribution ceases to change, for $t=21$ we got 
$\alpha=1.5$. 

The change in behavior for times larger than about $3$ (in units of the
divergence time) is connected with the discretness of the particles.  The
exponential contraction near the minima in the potential leads to a depletion of
particles in the neighborhood, so that the density cannot increase further by
accumulation once all particles that initially were in a region with negative
divergence are collected in a cell.  Estimates of the size of the cells can be
based on the spatial correlations of the divergence field, $C(r_i)=\langle
{\bf\nabla\cdot v}(r_{i0}+r_i) {\bf\nabla\cdot v}(r_{i0})\rangle$ with
$r_i=x,y$.  The typical extension of the patches, identified from the first zero
of the correlations, is about $20 \eta$ (inset of Fig.~\ref{fig4}).  Based
on this decorrelation length $l_d\approx 20 \eta$ the typical maximal
number of particles is about $n/N\sim l_d^2/L^2\sim 0.01$, in good agreement
with the data in Fig.~\ref{fig2}.

As a second set of characteristics we consider
the single-particle dispersion, $\sigma(t)$, and the two-particle pair 
(or relative) dispersion, $d(t)$. The first one is defined as the root 
mean square of
the absolute particle displacement, 
$\sigma(t)=\langle [{\bf x}(t; {\bf x}_0,0)-{\bf x}_0]^2\rangle_{\cal L}
^{1/2}$,
where $\langle\cdot\rangle_{\cal L}$ denotes an average over
the single Lagrangian particles. The second uses the difference 
${\bf R}_{12}(t)={\bf x}_1(t;{\bf x}_{1,0},0)- {\bf x}_2(t;{\bf x}_{2,0},0)$
between Lagrangian particle tracks that start at 
${\bf x}_{1,0}$ and ${\bf x}_{2,0}$ and is defined as
the root mean square value for all particles pairs,
$d(t)=\langle [{\bf R}_{12}(t)-{\bf R}_{12}(0)]^2\rangle_{\cal L}^{1/2}$.
In order to fix the dependence on initial separation we take it to be about
$1\eta$, with random orientation in space.  The single-particle dispersion
reflects the influence of flow structures at different scales on the particle
motion and the relative dispersion can detect the clustering, an interparticle
property.  In two-dimensional incompressible flows, the limiting cases for both
quantities are well-known \cite{Elhmaidi93}.  Both quantities have a ballistic
regime, $\sim t$, for short times when particle distances lie within the viscous
subrange.  For times much larger than the Lagrangian integral time scale, $T_L$,
correlations can be expected to have decayed, and the relative or single
particle motion becomes statistically independent and both dispersions increase
diffusively as in an uncorrelated Brownian motion, i.e.  $\sim t^{1/2}$.  For
intermediate times, anomalous scaling, $\sim t^{\beta}$ with $1/2<\beta<1$ has
been observed.  For an inverse Kolmogorov cascade, pair dispersion scaling
exponents were found to be close to the classical Richardson value of 3/2
\cite{Richardson} in numerical simulations \cite{Zouari94,Boffetta00} as well as
experiments \cite{Tabeling99}, but, e.g., sensitive to initial pair separation.
In the case of single particle dispersion transient trapping of tracers in
coherent vortex structures \cite{Elhmaidi93} affects the value of $\beta$.

\begin{figure}
\begin{center}
\epsfig{file=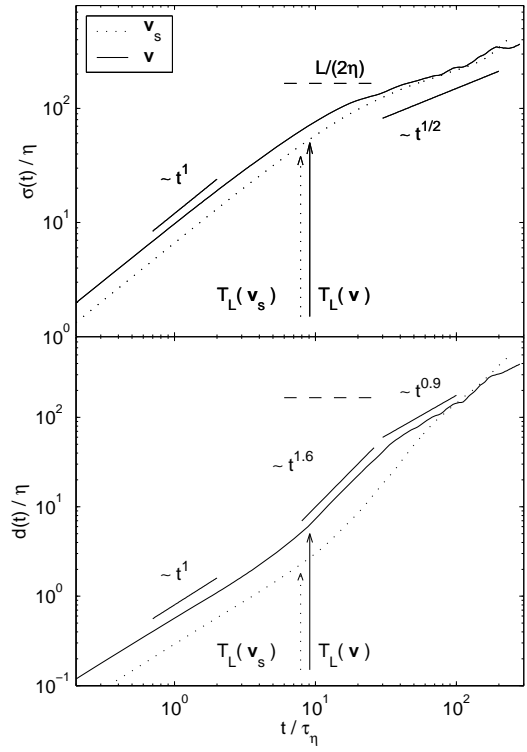,width=7cm}
\end{center}
\caption[]{Upper panel:  Single-particle dispersion, $\sigma(t)$, as a function
of time.  Solid curve is for advection by the full surface flow ${\bf v}$, while
the dotted one is for advection by the solenoidal part ${\bf v}_s$ only.  The
Lagrangian integral time scales for both fields are indicated by arrows.  Lower
panel:  Two-particle pair dispersion, $d(t)$, as a function of time.  Linestyles
are as above.  The dashed horizontal lines indicate half the box size.}
\label{fig5}
\end{figure}

Results on $\sigma(t)$ and $d(t)$ for the full surface flow and the solenoidal
part alone are shown in Fig.~\ref{fig5}.  The integral length scales are
$T_L/\tau_{\eta}=9.1$ for ${\bf v}$ and slightly shorter, $T_L/\tau_{\eta}=7.8$,
for ${\bf v}_s$.  In all cases we do observe the initial ballistic regime up to
$T_L$.  The single-particle dispersion crosses over to the Brownian regime,
$\sigma(t)\sim t^{1/2}$, for $t>T_L$ in both flow fields as indicated in the
upper panel of Fig.~\ref{fig5}.  We connect this behavior to the fastly varying
divergence patches \cite{Eck01} that cause a kind of stochastic sweeping of the
tracers.

For intermediate times pairs separate superdiffusively like $d(t)\sim t^\beta$
with an exponent of about 1.6, a value that is close to the Richardson
prediction, $d_R(t)\sim t^{3/2}$.  Small differences may also be attributed to
the additional fact that the surface flow was found to have larger intermittency
corrections than the associated three-dimensional volume turbulence
\cite{Gol01}.  Similar scaling behavior is observed for the pair advection in
the solenoidal part only (see the lower panel of Fig.~\ref{fig5} for both).

While this anomalous scaling continues for advection in ${\bf v}_s$ to even
larger times, we find a change of the pair dispersion to an almost
ballistic
behavior of $d(t)\sim t^{0.9}$ for $t\gtrsim 50\tau_{\eta}$, that does not seem
to cross over into a Brownian regime.  Microscopically, this means that while
one particle follows its partner within a pair, pair correlations decay more
slowly and $d(t)$ grows more rapidly than in the Brownian case.  Such almost
ballistic scaling was also found for the single-particle dispersion in the
strongly compressible one-dimensional Kuramoto-Sivashinsky equation \cite{Bohr}.
The difference to our case might be caused due to the dimensionality of the
problem and the specific character of our surface flow.  Superimposed on this
process is a chaotic component that comes from ${\bf v}_s$ and causes
exponential separation.  Another separation mechanism is a breaking of larger
particle clusters due the rapidly emerging flow, i.e.  due to the rapidly
changing divergence patterns.  Clustering and separation were found to be
competing processes that cause anomalous diffusion on a longer transient phase
of the evolution.

The exponential concentration described above has implications for the dynamics
of (inertial) particles that are not density matched with the fluid in which
they move.  The relations by Maxey and Riley \cite{Maxey} for their motion
implies that the velocity field of the particles is not divergence free.  The
particles will then cluster exponentially, as in Eq.~(\ref{psigmaeq}).  With a
view towards the formation of rain \cite{Balkovsky01} there is a uniform
condensation of droplets from thermodynamic nucleation and then an exponential
clustering to form larger drops, which then fall to the ground as rain drops.
The sizes of clusters thus range from the small scale droplets to the size of
rain drops and their size distribution thus reflects the distribution of
divergence fluctuations by Eq.~(\ref{psigmaeq}).

We thank J.~R.~Cressman, J.~Davoudi, W.~I.~Goldburg, E.~Hasco\"et, 
V.~Horvath,
K.~R.~Sreenivasan, and P.~K.~Yeung for discussions, and the John von
Neumann-Institut f\"ur Computing in J\"ulich for support and computing time on a
Cray T90.  One of us (J.S.)  would like to thank the Alexander von Humboldt
Foundation for support within the Feodor-Lynen program and Yale University for
support and hospitality.  This work was supported in part by the European
Community, HPRN-CT-2000-00162.

\narrowtext

\vfill
\end{multicols}
\end{document}